\documentclass[draft,11pt]{article}

\usepackage{amsmath}
\usepackage{latexsym}
\usepackage{amssymb}
\usepackage{bm}
\usepackage{graphicx}
\usepackage[all]{xy}
\newtheorem{defi}{***}
\newtheorem{lem}[defi]{Lemma}
\newtheorem{thm}[defi]{Theorem}

\def\tr{\mathop{\rm tr}\nolimits}\def\Tr{\mathop{\rm Tr}\nolimits}

\def\Id{I}

\def\rank{\mathop{\rm rank}\nolimits}

\def\argmin{\mathop{\rm argmin}}

\def\real{\mathbb{R}}
\def\complex{\mathbb{C}}

\def\Label#1{\label{#1}\ [\ #1\ ]\ }
\def\Bibitem#1{\bibitem{#1}\ [\ #1\ ]\ }
\def\Label{\label}
\def\Bibitem{\bibitem}

\begin{document}\setlength{\baselineskip}{18pt}

\bibliographystyle{unsrt}

\title{Quantum estimation and the quantum central limit theorem\footnote{The original Japanese version of this manuscript was published as an introductory article of quantum estimation in Bulletin of Mathematical Society of Japan, {\em Sugaku}, Vol.~55, No.~4, 368--391 (2003); Received September 2, 2002, 
Published October 24, 2003.
It was translated to English by Michikazu Sato and Masahito Hayashi.
The essential content of this manuscript is the same as that of the original version, while several descriptions and references are improved.
The material in this paper was presented in English in part at
{\em Joint MaPhySto and QUANTOP Workshop on Quantum Measurements and Quantum Stochastics}, 
Department of Mathematical Sciences, University of {\AA}rhus, Denmark, 2003,
{\em Non-locality of Quantum Mechanics and Statistical Inference}, 
Kyoto Sangyo Univ., Kyoto, 2003, and 
{\em Special Week on Quantum Statistics}, 
Isaac Newton Institute for Mathematical Sciences, Cambridge, UK, 2004.
The author was with the Laboratory for Mathematical Neuroscience, Brain Science Institute, RIKEN, Wako, Saitama, 351-0198, Japan. 
He is now with ERATO-SORST Quantum Computation and Information Project,
Japan Science and Technology Agency (JST), Bunkyo-ku, Tokyo, 113-0033, Japan, and 
Superrobust Computation Project,
Information Science and Technology Strategic Core (21st Century COE by MEXT) 
Graduate School of Information Science and Technology,
The University of Tokyo, Bunkyo-ku, Tokyo, Japan
(e-mail: masahito@qci.jst.go.jp).}}

\author{Masahito Hayashi}

\date{}
\maketitle

\section{Introduction}\Label{s1}
Recently, various quantum information-processing technologies have been proposed. Quantum statistical inference is included in these technologies. It is a feature of quantum information processing that any information is described by a small number of quanta (i.e., particles that behave in quantum mechanics). In traditional information processing that can be considered classical, any media communicating information can be considered to behave in classical mechanics: In such circumstances, from the viewpoint of quantum mechanics, any information can be described by quanta in order of the Avogadro number, so we can extract necessary information almost without any state reduction. However, we need to consider the state reduction by measurement in a microscopic world where we cannot apply the classical approximation, that is, any information of interest is not described by a sufficiently large number of quanta.

In such circumstances, it is required to extract information from obtained data efficiently. Generally, investigations on optimizations of data processing (e.g., estimating process) are called statistical inference. 
In particular, quantum systems have a problem due to measurement, which demolishes the state describing the information. Therefore, we can measure the system only once. (Any outcome of any repeated measurement can be described by an outcome of single measurement.)
For suitable estimation, we need to optimize our measurement. Furthermore, noncommutative physical quantities (observables) cannot be measured simultaneously. This fact makes the problem more difficult. In future, we can expect that this framework is applicable to nanotechnology and information processing systems that are excessively integrated. This formulation is called quantum estimation, which started in 1967 in relation to studies on optical communications from the quantum viewpoint.

On the other hand, uncertainty in quantum mechanics has attracted and fascinated many people. It is natural awareness of the issues to formulate the uncertainty, which is peculiar in quantum systems, and to aim to rigorous discussion. 
In such circumstances, quantum estimation can be regarded as a theme that formulates the uncertainty in quantum mechanics with mathematical rigor. Furthermore, as we will mention in \S\ref{s2}, owing to mathematical difficulty for quantum systems (or noncommutativity), quantum estimation requires higher mathematical theories that are not used in traditional statistics.
On one hand, quantum estimation is a mathematical theory based on experimental facility and statistical estimation, a kind of engineering framework; on the other hand, it is attractive as a basic problem of quantum mechanics and a new mathematical theory. Such multiplicity in quantum estimation is an obstacle for going on the study, while it is also rare attraction in this field.

Now, we will briefly pursue its development historically. In 1967, Helstrom~\cite{Helstrom:1967} started a study of quantum estimation introducing the symmetric logarithmic derivative (SLD)\null. 
In 1970s, excel studies are made: Yuen and Lax \cite{YL} studied estimation for a complex amplitude of coherent light in thermal noise and Holevo~\cite{HoG} studied estimation for the family of the quantum Gaussian states.
Furthermore, Holevo introduced the right logarithmic derivative \cite{HoG,HolP}, formalized group-covariant parametric estimation \cite{HoC,HolP}, and so on. In the beginning of 1980s, the studies in this field stagnated, but later in 1980s, the studies from a viewpoint different from those in 1970s began and developed rapidly \cite{HM2}. 
For example, we can list estimation in a two-parameter model of quantum two-level systems by Nagaoka~\cite{Nagaoka:1991}, 
estimation in a three-parameter model of quantum two-level systems by Hayashi~\cite{Haya1,Haya1-2}, 
and an approach to estimation theory in quantum two-level systems by evaluating the total Fisher information by Gill and Massar \cite{GM}. On the other hand, Fujiwara and Nagaoka \cite{FujiwaraNagaoka:1995,FujiwaraNagaoka:1996,FujiwaraNagaoka:1999}, Hayashi~\cite{Haya1-2,H-d}, and Matsumoto~\cite{Matsumoto,Matsumoto3} analyzed in detail the cases where it is known that the unknown state is a pure state. In addition, Hayashi~\cite{Haya97} applied irreducible representation theory of general linear groups \cite{Weyl} to statistical inference for quantum systems. Through detailed discussions, Matsumoto~\cite{Matu-Memo} derived the bound of estimation error for arbitrary finite-dimensional models. Furthermore, Nagaoka \cite{Nag,Na1} and Hayashi~\cite{Haya2002} discussed the state estimation from the viewpoint of large deviation, where two quantum versions of the Fisher information are compared and considered from the unified viewpoint. 
On the other hand, in the estimation of eigenvalues of the unknown density matrix, we have only one quantum version of Fisher information. Matsumoto~\cite{Matu-Memo} and Keyl and Werner \cite{KeylW} discussed this problem using the irreducible representation theory of general linear groups, which we mentioned before. The latter treated only large deviation, while the former treated also the mean-square error. In quantum estimation theory, it is a theme to improve the accuracy of estimation by using quantum correlation in the measurement apparatus. As the first study in this direction, Hayashi~\cite{Haya1-2} treated simultaneous estimation of the complex amplitude parameter and the photon number parameter in a quantum Gaussian state. Besides, Hayashi~\cite{bussei,qit8} showed that, use of correlation in measurement apparatus improves the accuracy of estimation of eigenvalues in the quantum two-level system under the assumption that any measurement is non-adaptive. Here, in an implementable framework, he discussed improvement using correlation in measurement apparatus over the case without correlation in measurement apparatus. 
In fact, the experiment proposed by the papers \cite{bussei,qit8} has been already implemented by Hendrich {\it et al.}\ \cite{HDFF}.
However, they concerned only the estimation by this method and did not concern the precision (MSE) of the estimation.
Furthermore, Fujiwara~\cite{Fuji01,Fuji02} applied the theory of estimation to a problem of estimating the quantum communication channel.

On the other hand, Kwiat {\it et al.}\ \cite{JKMW,KBSG, KBAW, KWWAE, WJEK} first made statistical estimation for the quantum state generated by an actual optical system. Usami {\it et al.}\ \cite{usami} demonstrated more accurate estimation based on fewer data generated by numerical experiment. 
In future, more accurate estimation would be implemented\footnote{Following these studies, Hayashi {\it et al.}\ established an optimization theory of measurement for testing entanglement \cite{HTM} and applied it to entangled photon pairs generated by SPDC \cite{HST}. Also the paper \cite{HTM} describes the experimental framework in more detail.}.

In this article, according to the organization mentioned below, we will treat the former studies from the unified viewpoint using the quantum central limit theory and so on. In \S\ref{s2}, we will outline the relation between quantum estimation and its related fields including the historical process. In \S\ref{s3}, we will summarize the formulation in quantum mechanics needed in minimum to treat quantum estimation. In \S\ref{s4} and \S\ref{s5}, we will explain the quantum Gaussian states, which can be regarded as the quantum version of Gaussian distributions, first for the one-mode case, second for the multi-mode case. Then, in \S\ref{s6}, we will describe the quantum central limit theorem and explain similarity and difference with the central limit theorem in classical systems. The part until \S\ref{s6} is the preliminary stage for the statistical discussion, and after this part we will have a particular discussion for statistical inference. In \S\ref{s7}, considering readers who are not familiar with classical (i.e., with respect to a family of probabilities) statistical inference, we will explain the formulation and the idea of statistical inference. 
There, we will explain that classical estimation can be regarded as a particular case of quantum systems mathematically: quantum estimation can be regarded as a generalization of classical estimation theory. Then, in \S\ref{s8}, we will summarize known results in classical systems. The following sections will treat their extension to the quantum case. First, in \S\ref{s9}, we will discuss estimation only for the expectation parameter of the family of the multi-mode quantum Gaussian states. Holevo \cite{HoG} mainly formulated this estimation, which can be treated relatively easily. In \S\ref{s10}, we will treat estimation theory for a general family of quantum states. Our methods of estimation will be divided into two classes by presence or absence of using quantum correlation in the measurement apparatus, and we will show that both bounds of estimation error do not coincide. We also will derive the general forms with respect to the bounds of estimation for both. In particular, in \S\ref{s11}, we will focus on the simultaneous estimation of the expectation parameter and the photon number parameter in the family of one-mode quantum Gaussian states. The proposed estimator will illustrate what quantum correlation in measurement apparatus is suitable for effective estimation. In \S\ref{s12}, we will describe the role of the quantum central limit theorem to derive the bound for the errors of estimators when quantum correlation is used in measurement apparatus. 
Finally, in \S\ref{s13}, we will describe themes in future based on these results. 

In this article, we will use somewhat complicated mathematical description. Operators on a Hilbert space corresponding to the quantum systems are denoted in capital letters, while constants are denoted in small letters. Furthermore, besides the Hilbert space, we often consider finite-dimensional vectors or matrices, which are denoted in small boldface, while their components are denoted in small lightface with a suffix (or suffixes, which we will not notice later). In addition, we sometimes consider vectors or matrices whose components are operators. They are denoted in capital boldface and their components, which are operators, are denoted in capital lightface with a suffix. When a vector or a matrix itself has a suffix, to denote its component, we write the suffix expressing its component first and after a semicolon (;) we write the suffix that the vector or the matrix itself has. For an $l$-dimensional vector or matrix $\bm{a}$, the vector or the matrix of the components from the first to the $d$th $(d\le l)$ is denoted as $\bm{a}_d$. The similar notation is used for vectors or matrices whose components are operators. Furthermore, for operators $X$ and $Y$, we write $[X,Y]:=XY-YX$ and $X \circ Y:=(XY+YX)/2$.

\section{Relation to other fields}\Label{s2}
Quantum estimation can be regarded as a kind of mathematical statistics of quantum systems. Main results in this field were not obtained with a mere combination of quantum mechanics and mathematical statistics, but they were obtained first with further detailed discussions. Here, referring its historical progress and its relation to quantum mechanics, we will explain its connection with related fields including mathematical statistics and quantum information theory.

Quantum theory, which describes a microscopic world, began in 1900 when Plank's constant was discovered. 
Then, it was formulated through the proposals of matrix mechanics by Heisenberg, wave mechanics by Schr\"{o}dinger, and probabilistic interpretation by Born. Furthermore, in 1930s, von Neumann succeeded in mathematically rigorous formulation with probabilistic interpretation and description of state reduction in measuring process \cite{von}. Probabilistic interpretation is essential for statistical inference in quantum systems because the estimation error and the error probability are concerned for this topic. However, von Neumann's formulation causes various problems because he adopted different principles in time evolution in a measuring process and usual time evolution. 
The conceptual problems of quantum mechanics, such as the measurement problem, seem to come from these double standards. 

In quantum systems, probabilistic interpretation is crucial when we measure one quantum (or a small number of quanta). In many actual physical experiments, however, observers measure quanta (e.g., atoms or molecules) with around Avogadro's constant (ca.\ $6 \times 10^{23}$) simultaneously, so they obtain only the average of the measured value directly. Owing to this reason, many studies based on quantum mechanics concern only the expectations of observables and rarely concern the probability distribution of measured data.
On the other hand, while there are many topics whose name contains ``quantum'' in mathematics, they focus only on higher mathematical difficulty such as infinity in quantum field theory, rather than problems related to probabilistic interpretation in quantum mechanics. Hence, statistical inference in quantum system is different from many traditional studies in that it targets not only the expectations of the measured data but their probability distributions themselves. In this mathematical stream with quantum mechanics, theory of Hilbert spaces was formulated and theory of operator algebra followed. Then, entropy theory \cite {OP} was expanded in relation to statistical mechanics, and furthermore it developed to quantum probability theory. Then, the quantum central limit theorem \cite {G-vW}, which is a theme in this article, was obtained by following these achievements, and it developed into free probability theory.

Here, we will briefly mention the relation between mathematical statistics and probability theory. Both fields treat the idea of probability, but it seems that they concern different aspects of probability. 
The former treats optimizations in some sense, e.g., the optimization of a random variable (an estimator etc.) for purposes of estimation and so on, while the latter takes an interest in mathematical structures. However, the central limit theorem has a basic position in both fields. In the former, it is widely accepted as a fundamental fact that the maximum likelihood estimator is asymptotically optimal and is asymptotically distributed as the normal distribution whose variance is the inverse of the Fisher information (or the Fisher information matrix) \cite{Lehmann,vdv}. This is essentially because the central limit theorem guarantees that the logarithmic differentiation of the probability density function is asymptotically distributed in the normal distribution as the number of data increases. Thus, the basic result of probability theory is effectively applied to mathematical statistics.

On the other hand, a quantization of random variables was done in quantum probability theory. However, its meaning is not clear from the viewpoint of simultaneous measurement of plural observables because it concerns not probabilistic structure but algebraic structure. Therefore, its result could not be applied to mathematical inference for quantum systems directly. At least, the quantum central limit theorem had not been considered applicable to evaluation of estimation error directly, as the classical one had. As we will describe in \S\ref{s12}, however, through the quantum central limit theorem, we can prove that a suitable estimator attains the asymptotically minimum error under the condition that quantum correlation is allowed in measuring apparatus.

While Matsumoto~\cite{Matu-Memo} obtained a similar statement for a general model in an imperfect form, our derivation is more natural and more general owing to the quantum central limit theorem.
In this discussion, the key is to construct an estimator in a somewhat technical form and to combine it with the quantum central limit theorem.

Quantum information theory is known as a field to discuss the quantum coding theorem and so on. In the latter 1960s, investigators in USSR began quantum information theory to find out how efficiently classical messages can be transmitted through a quantum communication channel (coding of a quantum communication channel). 
They include Holevo~\cite{Holevo-bounds,Holevo-bounds2}, Levitin~\cite{Lev}, and Stratonovich~\cite{Str}, who is one of the founders of stochastic differential equations. This problem consists of both optimizations of the encoding process and the decoding process. The latter optimization can be regarded as a decision problem of unknown quantum states, and Yuen, Kennedy, and Lax \cite{YKL} formulated theory of the optimum receiver for given signals. Thus, as is different from quantum probability theory, quantum information theory has developed since early time, related closely to mathematical statistics for quantum systems similar to quantum estimation. 
Holevo obtained an upper bound for the communication rate of the quantum communication channel in his two 1970s' papers 
\cite{Holevo-bounds,Holevo-bounds2}. After no less than twenty years, in the late 1990s, Holevo \cite{HoCh} and Schumacher and Westmoreland \cite{SW} independently proved that his upper bound can be attained. While this result requires the assumption that the communication channel be stationary memoryless, Hayashi and Nagaoka \cite{Hay-Nag} recently obtained a coding theorem for general quantum communication channels that are not necessarily stationary memoryless.
In the decoding process to achieve the optimal rate, so-called the square-root measurement is used in these results. 
Hayashi and Nagaoka \cite{Hay-Nag} introduced an important inequality for evaluation of the error probability of this decoding. Indeed, in quantum estimation, the square-root measurement is used for constructing an asymptotically optimal estimator, as we will mention in \S\ref{s12} of this article. 
Jointly with the quantum central limit theorem, this inequality plays an important part in the evaluation for the error of the proposed estimator.
Note that Holevo~\cite{Ho-kai} made detailed explanation on topics about the coding theorem of a quantum communication channel.

In the first half of 1990s, many researchers appeared in North America and Europe from related fields in quantum information theory.
Then, besides the coding theorem of a quantum communication channel mentioned above, the following problems began to be discussed: transmitting the quantum states themselves, compressing the quantum states, and so on. 
In particular, compressing the quantum state, which Schumacher began, is deeply related with quantum estimation, so we will explain it here. 
This protocol, which depends on the density operator of this ensemble, is said to be fixed-length because its rate of compression is fixed in advance.
Jozsa and the Horodecki family \cite{JH} proposed a fixed-length compression protocol depending only on the compression rate, which works for any stationary memoryless ensemble with the smaller von Neumann entropy than the compression rate. This protocol is called the universal fixed-length compression protocol.

Furthermore, Hayashi and Matsumoto \cite{HayaMa} proposed a compression protocol that works for any stationary memoryless ensemble. In such a protocol, the rate of compression should not be fixed in advance, i.e., a variable-length protocol is needed. In this compression protocol, a compression rate needs to be decided based on the input quantum state but then measurement is inevitable. If we assume only a stationary memoryless condition, the state demolition is inevitable, so it is impossible to decode a message without any error. Thus, in variable-length compression, we need to treat the tradeoff between state demolition and estimation of the compression rate. Asymptotically, however, we can estimate the compression rate as precisely as the optimal case with scarce state demolition. 
This protocol is called universal variable-length, and is constructed based on state estimation and Weyl's dual representation theory of permutation groups and general linear groups on a tensor-product space \cite{Weyl,GW}. 
As a commentary of quantum information theory including transmission of quantum states, see, for example, Hayashi~\cite{H-nyu}.

Next, we go on to another problem in quantum statistics inference, statistical hypothesis testing on quantum states, in which we assume two hypotheses for the unknown quantum state, and treat the tradeoff between two kinds of error probabilities.
Nagaoka focused on this problem when each hypothesis is of a single quantum state (i.e., a simple case), and proposed a quantum information-spectrum method~\cite{Nagaoka:1999}, which is the quantum version of information-spectrum methods~\cite{Han}. Testing two simple hypotheses is formulated by the Neymann--Pearson fundamental lemma in a very simple form with the non-asymptotic framework for classical cases. 
Holevo~\cite{Ho72} made its quantum extension in 1970s. In the classical case, Stein's lemma is known as fundamental fact that describes the asymptotic behavior of the optimal Type II error probability with the constant constraint for the Type I error probability.
Its quantum version was obtained by combining results of Hiai and Petz \cite{HP} and Ogawa and Nagaoka \cite{Oga-Nag:test}. Furthermore, Hayashi~\cite{H2001} characterized quantum measurement attaining the asymptotic optimal performance in the sense of Stein's lemma. This characterization was made with irreducible representation theory of general linear groups mentioned above. Interestingly, it depends only on one of two hypotheses. Ogawa and Hayashi \cite{Oga-Hay} analyzed the cases with an exponential constraint on one error probability. 
Furthermore, Ogawa~\cite{Oga} and Hayashi and Nagaoka \cite{Hay-Nag} found a close relation between the quantum version of Stein's lemma and the quantum channel coding theorem described before. Besides, using information-spectrum methods, Nagaoka and Hayashi \cite{Nag-Hay} formulate an infinite sequence of tests of general simple hypotheses. As a by-product of this formulation, they obtained theory in some extent to a general ensemble of quantum states that does not satisfy the stationary memoryless condition. 
In addition, extending theory of quantum hypothesis testing, Nagaoka formulated estimation theory based on large deviation type evaluation \cite{Nag,Na1}. Hayashi showed that two types of quantum Fisher information gave the bounds of the estimation error in two subtle different formulations \cite{Haya2002}.

Finally, we will describe the relation of quantum estimation with information geometry, which Amari and Nagaoka \cite{Amari,AN} formulated by regarding statistical estimation as geometry of a family of probability distributions. They introduced the idea of dual connection first, which plays an important role.
Nagaoka \cite{Nagaoka:1989:2,Nag94} started a quantum extension of information geometry, and Petz {\it et al.}\ \cite{Petz2} characterized the quantum version of Fisher information geometrically (for details, see Amari and Nagaoka \cite{AN}). Petz studied one type of quantum version of Fisher information, called Bogoliubov Fisher information, and showed that it is  the most natural among several quantum versions of Fisher information from the geometrical viewpoint. As Nagaoka~\cite{Nag,Na1} and Hayashi~\cite{Haya2002} discussed in detail, however, it turned out that such a geometrically natural Fisher information does not give a bound meaningful for estimation theory. Thus, it seems failed to extend the geometrical characterization of statistical estimation by dual connection to the quantum case, while this scenario might be expected in a naive sense by several researchers. 
Some quantum expansions of information geometry have been discussed only from the geometrical viewpoint, but it seems necessary to grasp them synthetically including the viewpoint of estimation in future. As a geometrical study including the viewpoint of estimation theory, interpreting Berry's phase~\cite{Shapere} as ``strength of noncommutativity," Matsumoto~\cite{Matsumoto} pointed out the relation between the bound of estimation error and Berry's phase in a family of pure states. Fujiwara \cite{fujiwara2,Fuj-kika} and Matsumoto \cite{Matsumoto2,Matsumoto3} developed this study to an internal relation between Berry's phase---and its extension, Uhlmann's parallelism \cite{Uhlmann:1986,Uhlmann:1993}---and Nagaoka's quantum information geometry.

\section{Quantum mechanics, especially states and measurement}\Label{s3}
To discuss statistical inference of quantum systems, it is necessary at least to describe the probabilities of observed values in measurement. However, though there are many textbooks about quantum mechanics, few of them describe the probabilities of observed values simply and rigorously. Though it is somewhat intrusive, we will explain the least items needed for description of probabilities of observed values in measurement. A theoretical description of state reduction by measurement has been already formulated rigorously, but we will not refer it here. See, e.g., Ozawa~\cite{ozawa}  or Hayashi~\cite{H-nyu}. 

First, a physical system of interest corresponds to a Hilbert space (a finite- or infinite-dimensional vector space with a Hermitian inner product), which is called its representation space. The system state is described with an operator $\rho$ called a density operator satisfying 
\begin{eqnarray}
 \rho=\rho^*,\quad \Tr\rho=1,\quad\rho\ge 0 .
\Label{defdensity}\end{eqnarray}
Furthermore, measurement on the system is described with a set of operators $M:=\{ M_\omega\}_{\omega\in\Omega}$ satisfying 
\begin{eqnarray}
 M_\omega= M_\omega^*, \quad M_\omega\ge 0,
 \quad \sum_{\omega\in\Omega} M_\omega= \Id,
\Label{defpovm}
\end{eqnarray}
where $M$ satisfying the above is called a {\it positive operator-valued measure (POVM)}. (Those who have read a standard textbook of quantum mechanics should note that this $M_\omega$ is not necessarily a projection.) Here, $\Id$ is the unit operator. We allow $\omega$ to take continuous values and $\sum$ signifies integration then. Unless it is confusing, we do not explicitly specify $\Omega$, in which $\omega$ takes values.

The density operator and the POVM defined above are mathematical representations of the state and the measurement of the system, respectively. Its concrete implication is given as follows: if we perform the measurement corresponding to $M=\{M_\omega\}_{\omega\in\Omega}$ to the state corresponding to $\rho$, then the probability to get $\omega$ is 
\begin{eqnarray}
{\rm P}^M_{\rho}(\omega):=
\Tr \rho M_\omega. \Label{a6}
\end{eqnarray} 
In this formulation, reproducibility of both the state and measurement is implicitly assumed 
(otherwise, it may be almost impossible to verify the formula~($\ref{a6}$) experimentally). 
We will call the system ${\cal H}$, the state $\rho$, the measurement $M$ for short.

According to the convention of physics,  when we regard each an element $u \in {\cal H}$ as an element of the dual space ${\cal H}^*$ through the inner product, we will denote it by $\langle u|$. When we want to emphasize that $u$ is not an element of ${\cal H}^*$ but is an element of ${\cal H}$, we may write $| u \rangle$. In other words, when we regard $|u\rangle=(u_1\,u_2\,\cdots)^T \in {\cal H}$ as a column vector, the transposed vector with its complex conjugate elements is denoted as $\langle u|=(u_1^*\,u_2^*\,\cdots)$. 
In the following, we denote the set of density matrices by ${\cal S}({\cal H})$, and we will explain its structure. If the system state is $\rho_1$ with probability $\lambda$ and $\rho_2$ with probability $1-\lambda$, by performing certain measurement $M=\{M_\omega\}$ to the system, the probability to get an observed value $\omega$ is 
\begin{eqnarray} 
\lambda \Tr \rho_1 M_\omega+(1-\lambda)\Tr \rho_2 M_\omega =\Tr\{(\lambda \rho_1+(1-\lambda) \rho_2) M_\omega \} \label{k1}. \end{eqnarray} 
Here, even if we regard the state of this system as $\rho'=\lambda\rho_1+ (1-\lambda)\rho_2$ and calculate the probability distribution of the data according to $(\ref{a6})$, no problem occurs on integrity with all theoretically possible experiments. Therefore, the state of the system can be regarded as $\rho'$. This is called probabilistic mixture (incoherent superposition) and is distinct from so-called quantum superposition. Furthermore, when the system is the composite system of systems ${\cal H}_1$ and ${\cal H}_2$, then the representation space is given by the tensor-product space ${\cal H}_1 \otimes {\cal H}_2$. 

For the discussion above, we may disregard the existence of the representation space and need only treat a $*$-algebra, which is a generalization of the linear space of operators as follows: A complex linear space $V$ is called a $*$-algebra when it has the unit element $I$, a product satisfying the linearity and the associativity, and a $*$-operation satisfying the following condition: $(c x)^*= c^* x^*$ for any $x\in V$ and any complex number $c$. 
For such a $*$-algebra, using the set of positive elements $\{x x^* | x \in V\}$, we can reconstruct the description of quantum systems mentioned above as follows: 
First, a measurement is given by a decomposition of the unit element $I$ by positive elements, that is, a generalized POVM $M:= \{M_\omega\}_{\omega}$. A state is given by an element $\rho$ of the dual space $V^*$ of $V$ satisfying $\rho(I)= 1$ and $\rho(xx^*) \ge 0$, that is, a generalized state. Then, when we perform the measurement $M$ to the system in the state $\rho$, the probability to get an observed value $\omega$ is given by $\rho(M_\omega)$. Note that, if $x^*=x$ holds, $e^{x}= e^{x/2}(e^{x/2})^* \in \{x x^* | x \in V\}$ because $(e^{x/2})^*=e^{x/2}$. 
Of course, if $V$ is a space of linear operators on ${\cal H}$, then the POVMs (the states) defined by (\ref{defpovm}) ((\ref{defdensity})) coincide with the generalized POVMs (the generalized states) given here, respectively. Thus, we can regard the $*$-algebra $V$ as a quantum system. Furthermore, the composite system of $*$-algebras $V_1$ and $V_2$ can be described with a $*$-algebra on the tensor-product space $V_1 \otimes V_2$ in which the following product and the $*$-operation: 
\begin{align} (x_1\otimes y_1 ) \cdot (x_2\otimes y_2 ) := (x_1 \cdot x_2) \otimes (y_1 \cdot y_2) , \quad
(x_1\otimes y_1 )^* := x_1^* \otimes y_1^*.
\end{align}
When $V$ is the set of linear operators on the Hilbert space, this tensor-product of $*$-algebras $V_1$ and $V_2$ coincides with the set of linear operators on the composite system defined in the former.

\section{One-mode quantum Gaussian systems}\Label{s4}
As a concrete example of quantum systems, we consider a Hilbert space $L^2(\real)$, which is often called the Fock space. By denoting by $|k \rangle$ the $k$th Hermitian function, $\{ |0\rangle, \ldots \}$ forms its orthonormal basis. This space describes the physical system of photons with a specific wavelength, and the state $|k \rangle \langle k |$ represents the state of $k$ photons. 

In this system, operators $Q$ and $P$ are defined by
\begin{align}
(Q f) (x):= x f(x) ,\quad
(P f) (x):= - i \frac{\,d f}{\,d x}(x).
\end{align}
These operators play an important role. They are called the position operator and the momentum operator when the quantum system represents a one-dimensional motion of a particle. 
They satisfy the commutation relation
\begin{align}
[Q,P]= iI. \Label{991}
\end{align}
By defining a coherent vector $|\alpha\rangle_a:=
\sum_{k=0}^\infty e^{-\frac{|\alpha|^2}{2}}\frac{\alpha^k}{\sqrt{k!}}
|k \rangle$ $( \alpha \in \complex)$,
they have the following relation with the annihilation operator 
$a:= \frac{1}{\sqrt{2}}(Q+ i P)$:
\begin{align}
a| \alpha\rangle_a= \alpha | \alpha\rangle_a.\end{align}
In addition, the density operator $\rho_{\alpha,0}:=|\alpha\rangle_a ~_a\langle \alpha|$ is called a coherent state. 
Two POVMs
\begin{align}
\xymatrix{*+[F-]{{\bf N}}}:&\quad k \mapsto 
| k \rangle \langle k| 
\Label{9}\\
\xymatrix{*+[F-]{{\bf H}}}:& \quad\alpha \mapsto 
|\alpha\rangle_a ~_a\langle \alpha| 
\Label{10}
\end{align}
are known as important measurements, and are called the number measurement and the heterodyne measurement, respectively. The former takes integer values, which are discrete; the latter takes complex values, which are continuous. We can show that $\xymatrix{*+[F-]{{\bf H}}}$ is a POVM by the equality 
\begin{align}
\int_\complex |\alpha\rangle_a ~_a\langle \alpha| \,d x
= I. \Label{995}
\end{align}

A coherent state is a relatively stable state and plays an important part in quantum optics. In particular, in this article, we will note a quantum Gaussian state that is a probabilistic mixture of coherent states by Gaussian integral $\rho_{\zeta,N}:=\frac{1}{\pi N}\int_{\complex} e^{-\frac{|\alpha-\zeta|^2}{N}}| \alpha\rangle_a ~_a\langle \alpha| \,d \alpha$. The quantum Gaussian state $\rho_{\zeta,N}$ can be described with $Q$ and $P$ as follows:
\begin{align}
\rho_{\zeta,N}=
\frac{\exp(\log (\frac{N}{N+1}) (a^*- \overline{\zeta})(a- \zeta))}
{N+1}
=
\frac{\exp\left(\log (\frac{N}{N+1}) 
\frac{(Q-\sqrt{2}\zeta_x)^2 + (P-\sqrt{2}\zeta_y)^2}{2} \right)}{N+1},
\Label{999}
\end{align}
where $\zeta= \zeta_x + i \zeta_y$. 
Furthermore, it is known that the quantum Gaussian state $\rho_{\zeta,N}$ satisfies the equation
\begin{align}
\Tr \rho_{\zeta,N}\exp(i(xQ +yP))
= \exp\left(i(\sqrt{2}\zeta_x x +\sqrt{2}\zeta_y y)
- \frac{1}{2}\left(N+\frac{1}{2}\right)(x^2+y^2)
\right)
,\quad \forall x, y \in \real. \Label{992}
\end{align}
Conversely, it is also known that a density operator satisfying (\ref{992}) is restricted to a quantum Gaussian state $\rho_{\zeta,N}$. Therefore, we can also regard (\ref{992}) as the definition of $\rho_{\zeta,N}$.

It is known that any operator of this system can be given as the limit of sums of algebraic terms of $Q$ and $P$ \cite{HolP}. Therefore, we can treat the quantum system described by the Hilbert space $L^2(\real)$ based only on the operators $Q$ and $P$ satisfying the commutation relation~(\ref{991}) without considering the structure of the original Hilbert space $L^2(\real)$.

\section{Multi-mode quantum Gaussian systems}\Label{s5}
The Fock space above describes a physical system corresponding with one wavelength. A quantum system of photons of plural $l$ wavelengths is described by a Hilbert space $L^2(\real^m)=L^2(\real)^{\otimes m} $. That is, this quantum system can be described by the space of operators generated by a sequence of self-adjoint operators $Q^1, \ldots Q^m, P^1, \ldots, P^m$ satisfying
\begin{align}
[P^k,P^j]= [Q^k,Q^j]= 0 ,\quad
[Q^k,P^j]= i s_k\delta^{k,j}I. \Label{20-8}
\end{align}

We will more generally consider a system defined by a sequence of self-adjoint operators satisfying 
\begin{align}
\frac{1}{2}[X^k,X^j]= is^{k,j}I \Label{812-a}
\end{align}
($1\le k,j\le d$). 
Of course, $\bm{s}=[s^{k,j}]$ is an antisymmetric matrix. 
As generalization of (\ref{992}), for a vector $\bm{\theta}= [\theta^k]$ and a symmetric matrix $\bm{v}=[v^{k,j}]$ satisfying $\bm{v} \ge 0$ and $\bm{v} + i \bm{s} \ge 0$, we define a quantum Gaussian state $\rho_{\bm{\theta},\bm{v}}$ as the state satisfying
\begin{align}
\Tr \rho_{\bm{\theta},\bm{v}} \exp\left(i \sum_k \eta_k X_k \right)
=
\exp\left(i \sum_k \eta_k \theta^k - \frac{1}{2} 
\sum_{j,k} v^{k,j} \eta_k \eta_j\right).
\end{align}
It is known that the condition above is equivalent to the following \cite{CCR}:
\begin{align}
\Tr \rho_{\bm{\theta},\bm{v}} X^k &= \theta^k, \Label{812-b}\\
\Tr \rho_{\bm{\theta},\bm{v}}((X^{k_1} - \theta^{k_1})
(X^{k_2} - \theta^{k_2}) \cdots (X^{k_l} - \theta^{k_l})) &=
\left\{
\begin{array}{cl}
0 & l= 2n+1, \\
\sum \prod_{h=1}^{n}
(v^{k_h,j_h} + i s^{k_h,j_h}) &l= 2n.
\end{array}\right.\Label{812}
\end{align}
In particular, for $l=2$ in (\ref{812}), the right-hand side is equal to $v^{k,j}$. Note that $\sum$ signifies the sum with respect to all kinds of decomposition $\{H_1, \ldots, H_n \}$ of $\{1, 2, \ldots, 2n\}$ of the form $H_h=\{k_h, j_h \}$, where $j_h < k_h$.

The existence of $\bm{X}$ and $\rho_{\bm{\theta},\bm{v}}$ can be shown as follows: First, assume that $d$ is even and the rank of $\bm{s}$ is $d$. Then, we select an adequate orthogonal matrix so that $\bm{a}\bm{v}\bm{a}^*$ is the unit matrix and $\bm{a}\bm{s}\bm{a}^*$ satisfies (\ref{20-8}), where $\bm{a}:= \bm{o}\bm{v}^{-1/2}$. Therefore, we can construct the sequence $\bm{X}$ of operators (the state $\rho_{\bm{\theta},\bm{v}}$) as operators (a state) on the representation space $L^2(\real^{d/2})$, respectively. Next, we consider the case when the rank of $\bm{s}$ is less than $d$ (for an odd $d$, this holds by antisymmetry). If the rank of $\bm{v}$ is $d$, adding adequate operators $X^{d+1}, \ldots, X^{2d- \rank \bm{s}}$, we can reduce this case to the former one. If the rank of $\bm{v}$ is also less than $d$, we change the coordinate adequately so that $\bm{v}$ and $\bm{s}$ have non-zero components at only the first $\rank \bm{v}$ components. Then, we can regard $X^{\rank \bm{v}+1}, \ldots, X^d$ as the constant $0$ and the rest elements $X^1, \ldots, X^{\rank \bm{v}}$ can be reduced to the former argument. In the following, we will disregard the structure of Hilbert space and discuss only operators characterized by (\ref{812-a}), (\ref{812-b}), and (\ref{812}).

Here, we define the operator $T_{\bm{\theta}',\bm{v}'}$ for a vector $\bm{X}$ consisting of operators as follows: \begin{align}
T_{\bm{\theta'},\bm{v'}}(\bm{X}):=
\frac{
\exp\left(
- (\bm{X}- \bm{\theta}')^* (\bm{v'})^{-\frac{1}{2}}
f ( |(\bm{v'})^{-\frac{1}{2}} \bm{s} (\bm{v'})^{-\frac{1}{2}}| )
(\bm{v'})^{-\frac{1}{2}} (\bm{X}- \bm{\theta}') \right)}
{(2 \pi)^{\frac{d}{2}} (\det(\bm{v}'))^{\frac{1}{2}}\det(1- |(\bm{v}')^{-\frac{1}{2}} s (\bm{v}')^{-\frac{1}{2}}|^{\frac{1}{4}})},
\Label{4-25-1}
\end{align}
where the function $f$ on $[0,1)$ is
\begin{align}
f(s):= \frac{1}{4s} \log \left(\frac{1+s}{1-s}\right)
,\quad 
f(0):= \frac{1}{2}.
\end{align}
Here, when the variable of $f$ is a matrix, $f$ is defined as a matrix function, i.e., the function $f$ is applied to the diagonal elements. Since $f$ is defined on $[0,1)$, we need to select $\bm{v'}$ so that all eigenvalues of $ |(\bm{v'})^{-\frac{1}{2}} \bm{s} (\bm{v'})^{-\frac{1}{2}}|$ are less than~$1$.

Then, for a multi-mode quantum Gaussian system, we can prove
\begin{align}
\int_{\real^{d}} T_{\bm{\theta}', \bm{v}'}(\bm{X}) \,d \bm{\theta}' & 
= I, \Label{812-1}\\
\Tr T_{\bm{\theta}', \bm{v}'}(\bm{X}) \rho_{\bm{\theta},\bm{v}} &=
\frac{\exp\left( - \frac{(\bm{\theta}- \bm{\theta}')^* 
( \bm{v}+ \bm{v}')^{-1}(\bm{\theta}- \bm{\theta}')}{2} \right)}
{(2 \pi)^{\frac{d}{2}} (\det(\bm{v}+\bm{v}'))^{\frac{1}{2}}}.
\Label{822}
\end{align}
Therefore, (\ref{4-25-1}) and (\ref{812-1}) show that $T_{\bm{\theta'},\bm{v'}}(\bm{X})$ gives a generalized POVM\null. Furthermore, for a sequence $\bm{X}_{d'}$ of the $d$ operators $X^1, \ldots, X^{d'}$, a $d'$-dimensional matrix $\bm{v'}$, and a $d'$-dimensional vector $\bm{\theta'}$, we define $T_{\bm{\theta'},\bm{v'}}(\bm{X}_{d'})$, similarly to the above, by
\begin{align}
T_{\bm{\theta'},\bm{v'}}(\bm{X}_{d'}) 
:=
\frac{
\exp\left(- (\bm{X}_{d'}- \bm{\theta}')^* (\bm{v'})^{-\frac{1}{2}}
f ( |(\bm{v'})^{-\frac{1}{2}} \bm{s} (\bm{v'})^{-\frac{1}{2}}| )
(\bm{v'})^{-\frac{1}{2}} (\bm{X}_{d'}- \bm{\theta}') \right)}{(2 \pi)^{\frac{{d'}}{2}} (\det(\bm{v}'))^{\frac{1}{2}}
\det(1- |(\bm{v}')^{-\frac{1}{2}} 
\bm{s}_{d'} (\bm{v}')^{-\frac{1}{2}}|^{\frac{1}{4}})}.
\Label{4-25-2}\end{align}
Then, similar equalities to (\ref{812-1}) and (\ref{822}) hold and 
$T_{\bm{\theta'},\bm{v'}}(\bm{X}_{d'}) $ is thought to give a generalized POVM.

\section{The quantum central limit theorem}\Label{s6}
For a density matrix $\rho$ on ${\cal H}$, we choose a self-adjoint operator $X$ satisfying $\Tr \rho X=0$, and denote by $X^{(n)}$, the self-adjoint operator $\frac{1}{\sqrt{n}}\sum_{j=1}^n X_{(j)}$ on ${\cal H}^{\otimes n}$, where $X_{(j)}$ means $I \otimes \cdots \otimes I \otimes X \otimes I \otimes \cdots \otimes I$. Here, consider a vector consisting of self-adjoint operators $\bm{X}=(X^1, \ldots, X^l)$ satisfying $\Tr \rho X^k =0$. Then, matrices $v^{k,j}_{\rho}(\bm{X}):= \Tr \rho (X^k \circ X^j)$ and $s^{k,j}_{\rho}(\bm{X}):=\frac{-i}{2} \Tr \rho [X^k,X^j]$ satisfy $\bm{v}_{\rho}(\bm{X}) + i\bm{s}_{\rho}(\bm{X}) \ge 0$. Concerning $X^{k,(n)}$ and a density matrix $\rho^{\otimes n}:= \underbrace{\rho \otimes \cdots \otimes \rho}_{n} $ on ${\cal H}^{\otimes n}$, the following theorem holds:

\begin{thm}\Label{t1}
For $\bm{X}$, we define a vector $\overline{\bm{X}}$ consisting of operators on a multi-mode Gaussian system satisfying $[\overline{X}^k, \overline{X}^j]= i s^{k,j}_{\rho}$. Then, the following equality holds:
\begin{align}
\lim_{n \to \infty} \Tr \rho^{\otimes n}P (\bm{X}^{(n)}) =
\Tr \rho_{0,\bm{v}_{\rho}} P(\overline{\bm{X}}),
\end{align}
where $P$ is an arbitrary polynomial and $\rho_{0,\bm{v}_{\rho}} $ is a quantum Gaussian state on $\overline{\bm{X}}$.
\end{thm}

This theorem is called the quantum central limit theorem (or the algebraic central limit theorem), which Giri and von Waldenfels \cite{G-vW} proved first, and Petz~\cite{CCR} serves as its introductory references. 
For an operator $Y$, we will denote by $\overline{Y}$ the corresponding operator on the multi-mode Gaussian system in the sense of the central limit theorem above. In this theorem, we regard a sequence of operators $X_1, \ldots ,X_d$ as a quantum version of sequence of random variables, like the classical case. Then, the expectation of their product converges to the expectation of the product for the Gaussian systems. However, it does not show that the operators $X_1, \ldots ,X_d$ are asymptotically commutative in some sense, so it is difficult to find out what kind of measurement is available in the asymptotic setting only from this theorem.

On the other hand, the equality (\ref{812-1}) guarantees that the operator $T_{\bm{\theta}',\bm{v}'}(\overline{\bm{X}})$ can be regarded as a measurement in the multi-mode Gaussian system. 
Since the operator $T_{\bm{\theta}',\bm{v}'}(\bm{X}^{(n)})$ can be characterized as the limit of the polynomial of operators $\bm{X}^{(n)}$, using Theorem~\ref{t1} and adequate discussion of analytic prolongation, we can show the following theorem\cite{H-pre}:
\begin{thm}\Label{t2}
If $\Tr \rho \exp(t (X^k)^2)\,< \infty$ for $i= 1, \ldots , d\le l$ and sufficiently small $t\,>0$, then the following equalities hold:
\begin{align}
\lim_{n \to \infty} \Tr \rho^{\otimes n} 
T_{\bm{\theta}_d',\bm{v}'_d}(\bm{X}^{(n)}_d)
&= \Tr \rho_{0,\bm{v}_d} T_{\bm{\theta}'_d,\bm{v}'_d}(\overline{\bm{X}}_d), \\
\lim_{n \to \infty} \Tr \rho^{\otimes n} X^{j;(n)} 
T_{\bm{\theta}',\bm{v}'}(\bm{X}^{(n)}_d)
&= \Tr \rho_{0,\bm{v}_d} \overline{X}^j T_{\bm{\theta}'_d,\bm{v}'_d}
(\overline{\bm{X}}_d).
\end{align}
\end{thm}
Hence, we can expect to apply these discussions to estimation theory where measurement is concerned. 

\section{Statistical inference in quantum systems}\Label{s7}
When a quantum system is generated from a new particle generator or we lack a part of information describing the system, the density matrix of the system is not {\it a priori}\/ known, so we infer the density matrix from the data of an experiment designed adequately.  
In particular, estimating the unknown density matrix is called {\it quantum estimation}. When we statistically infer the density matrix of the system, we need to prepare plural systems of the same state. For example, when the particles are repeatedly generated by a certain particle generator under the same condition, the states of these systems may be regarded almost as the same states.

Under such a framework, from the prior information, we often assume that the unknown density matrix belongs to a certain {\it family of density operators (matrices)}\/ ${\cal S}= \{ \rho_{\bm{\theta}} \in {\cal S}({\cal H}) | \bm{\theta} \in \Theta \subset \real^d \}$. Then, estimating the density matrices are equivalent with estimating the unknown parameter from observed data.

For example, when ${\cal H}= \complex^2$, unless we have prior knowledge, we need to estimate the parameter $(x,y,z)$ in the family of density matrices
\begin{eqnarray*}{\cal S}= 
\left\{ \left.\rho_{x,y,z}=
\frac{1}{2}\left( 
\begin{array}{cc}
1+x & y +iz \\
y - iz & 1-x 
\end{array}
\right) 
\right| 1 \ge x^2+y^2 +z^2 \ge 0\right\}.
\end{eqnarray*}
For the same system, if it is known (or can be considered) that $\Tr \rho \left(\begin{array}{cc}0 & i \\ -i & 0 \end{array}
\right)=0$, then we can assume that the unknown density matrix belongs to the family of density matrices
\begin{eqnarray*}
\displaystyle
{\cal S}_{z=0}:=
\left\{ \left. \rho_{x,y,0}=
\frac{1}{2}\left( 
\begin{array}{cc}1+x & y \\
y & 1-x 
\end{array}
\right) 
\right|1 \ge x^2+y^2 \ge 0\right\}.
\end{eqnarray*}
Therefore, we need only estimate the unknown parameter $(x,y)$ from the observed data.

As another example, consider a system of photons with the frequency $f$. 
This system is considered to develop as time evolution according to the following Master equation \cite{SZ}:
\begin{align*}
\frac{\,d \rho}{\,d t}
= -i\left[ \nu\left(a^*a+\frac{1}{2}\right), \rho\right]
-\frac{c\overline{n}}{2}(aa^*\rho-2 a^*\rho a + \rho a a^*) 
- \frac{c(\overline{n}+1)}{2}
(a a^* \rho - 2 a \rho a^* + \rho a^* a),\end{align*}
where $c$ is the coupling constant with the environment system 
and $\overline{n}$ is the average photon number in the environment system. Thus, the coherent state $| \zeta_0\rangle_a ~_a\langle \zeta_0| $ evolves in time as 
$| \zeta_0\rangle_a ~_a\langle \zeta_0|\mapsto
\rho_{\zeta_0 e^{-\frac{ct}{2}-i \nu t},
\overline{n}(1-e^{-ct})}$.
Since the coherent state is natural for the initial state, it is natural to assume that the unknown final state belongs to the one-mode quantum Gaussian states family. 
If $N$ is known, then the problem is estimation of the two-dimensional parameter $(\zeta_1,\zeta_2)$ in the family of density operators ${\cal S}_N^g:= \{ \rho_{\zeta,N}| \zeta (= \zeta_1 + \zeta_2 i )\in \complex \}$. If $N$ is unknown, then the problem is estimation of the three-dimensional parameter $(\zeta_1,\zeta_2,N)$ in the family of density operators ${\cal S}^g:= \{ \rho_{\zeta,N}| \zeta \in \complex , N \,> 0\}$.

More generally, if the system is ${\cal H}$, measurement is described by a POVM $M$ on ${\cal H}$. In particular, when we estimate an unknown state that belongs to a family of density operators 
${\cal S}=
\{ \rho_{\bm{\theta}} \in {\cal S}({\cal H}) |
\bm{\theta} \in \Theta \subset \real^d \}$, 
we need a mapping from the set $\Omega$ of observed values to the parameter space $\Theta$ besides a POVM. 
The joint process of this measurement and this mapping can be described by a POVM that takes values in the parameter space $\Theta$, which is called {\it an estimator}. We often discuss statistical inference under the assumption that $n$ unknown states identical to $\rho_{\bm{\theta}}$ are prepared independently. Then, the total system is written by the Hilbert space ${\cal H}^{\otimes n}$, and its density operator is $\rho_{\bm{\theta}}^{\otimes n} $. 
Hence, we treat estimation concerning the family of density operators $\{\rho_{\bm{\theta}}^{\otimes n} \in {\cal S}({\cal H}) |
\bm{\theta} \in \Theta \subset \real^d\}$, 
and any estimator is given by a POVM on ${\cal H}^{\otimes n}$.

In statistical inference of quantum systems, now, we assume that one measurement $M$ is chosen and cannot be changed.
Then, estimation of the density matrix can be reduced to estimation in a family of distributions $\{P^M_{\rho_{\bm{\theta}}}|\bm{\theta} \in\Theta\}$. 
To investigate discussion peculiar to quantum system, we have to discuss optimization of measurement $M$. 
This can be also considered a kind of experimental design.

For example, when the unknown state is $\rho_{\bm{\theta}}$ and an estimator $M$ is applied, the $k$th element $e^{k}_{\bm{\theta}}(M)$ of the mean of the estimate $\bm{e}_{\bm{\theta}}(M)$ is equal to
\begin{align}
e^k_{\bm{\theta}}(M)=
\int_{\Theta} \hat{\theta}^k \Tr \rho_{\bm{\theta}} M(\,d \hat{\bm{\theta}})
\end{align}
and the mean-square error matrix $\bm{v}_{\bm{\theta}}(M)=[v^{k,j}_{\bm{\theta}}(M)]$ becomes
\begin{align}
v^{k,j}_{\bm{\theta}}(M)=
\int_{\Theta} (\hat{\theta}^k- \theta^k)(\hat{\theta}^j- \theta^j)
\Tr \rho_{\bm{\theta}} M(\,d \hat{\bm{\theta}}).\end{align}
In statistical inference, we mainly evaluate errors with the mean-square error matrix. We often pose the condition of unbiasedness 
\begin{align}
\bm{e}_{\bm{\theta}}(M)= \bm{\theta}
\end{align}
to estimators. 
This condition, however, depends on the choice of the coordinates. 
Therefore, we need not always restrict estimators within this condition for discussion.

When all density operators in the given family are commutative, they can be diagonalized simultaneously, and the optimal measurement is given as the resolution of the unity based on the orthonormal system consisting of common eigenvectors. 
Here, label the common eigenvectors of these density matrices by $\omega=1,2,\ldots$. Then, diagonal elements $p_{\bm{\theta}}(\omega)$ form a probability distribution. We can reduce our problem to estimation of the unknown parameter in a family of probability distributions $\{p_{\bm{\theta}} = \{ p_{\bm{\theta}}(\omega)\}| \bm{\theta} \in \Theta \subset \real^d\}$. When the system follows classical mechanics, all states are described by probability distributions, so the estimation problem concerning a family of probability distributions is said to be classical. 
Even in such classical cases, it is generally difficult to minimize the mean-square error matrix $\bm{v}_{\bm{\theta}}(M)$ for all $\theta$. Therefore, for classical models, we often discuss only the cases of sufficiently large $n$ (asymptotic cases).

Here, we need to explain briefly why the composite system of two same quantum systems is represented not by the symmetric or antisymmetric tensor-product space, but by the tensor product system ${\cal H}^{\otimes 2}$. Generally, even in the ``same'' experiments in classical setting, their places and their times are different. That is, even if the two quantum systems can be regarded as the same, usually only the microscopic parts are the same and represented by ${\cal H}$, but other parts (e.g., the position and the time) are different. 
Hence, if the states of these different parts are known and if we focus only on the same system, then the total focused system cannot be treated by Bose statistics or Fermi statistics and the representation space is the tensor product space ${\cal H}^{\otimes 2}$. 
This is because these two particles can be distinguished by the additional parts (e.g., the position). For example, $n$ spin 1/2 particles have the same system $\complex^2$ concerning the spin, but to treat their position, we need another representation space. 
If we know the position of each particle, and if we focus only on their spin, the system of the total spin is represented by the $n$-tensor product space of $\complex^2$ because we identify each particle. Here, it is assumed that plural identical systems are prepared, not that we make copies of the same state. It is impossible to clone a quantum state perfectly.

As we wrote in \S\ref{s1}, if the number of particles is quite huge, the average of this ensemble obeys classical mechanics. This number is quite different from the number required for the application of asymptotic theory of statistical inference. A number large for statistical inference is ten thousand or a hundred thousand. On the other hand, it is considered that approximation of classical mechanics can be applied when the number is about Avogadro's constant (ca.\ $6 \times 10^{23}$) or its square root. Therefore, even if asymptotic theory can be applied, we still need to use the quantum framework.

\section{Asymptotic theory in classical systems}\Label{s8}
We will briefly explain asymptotic theory in classical systems. First, for a family of probability distributions $\{p_{\bm{\theta}} = \{ p_{\bm{\theta}}(\omega)\}| 
\bm{\theta} \in \Theta \subset \real^d\}$, we define the Fisher information matrix $\bm{j}_{\bm{\theta}}=[j_{k,l;\bm{\theta}}]$ by
\begin{align}
j_{k,l;\bm{\theta}} := \sum_\omega p_{\bm{\theta}}(\omega) 
\frac{\partial \log p_{\bm{\theta}}(\omega)}{\partial \theta^k}
\frac{\partial \log p_{\bm{\theta}}(\omega)}{\partial \theta^l}.
\end{align}
Then, for an unbiased estimator $M$, the following Cram\'{e}r--Rao inequality holds:
\begin{align}
\bm{v}_{\bm{\theta}}(M) \ge \bm{j}_{\bm{\theta}}^{-1}. \Label{Cr}
\end{align}
In the proof of the inequality (\ref{Cr}), the Schwarz inequality plays an essential role. For details, see \cite{Amari,AN}. 
As we mentioned before, however, we need not restrict estimators within unbiased ones but we need to discuss estimators in a wider class. 
When the number $n$ of the prepared states is sufficiently large, it is relatively easy to treat the problem: If $n$ is large, for an estimator $M^n$, 
the mean-square error matrix $\bm{v}_{\bm{\theta}}(M^n)$ asymptotically satisfies the inequality
\begin{align}
\bm{v}_{\bm{\theta}}(M^n) \gtrsim \frac{1}{n}\bm{j}_{\bm{\theta}}^{-1} .
\Label{kka}
\end{align}

Furthermore, the maximum likelihood estimator $\bm{\theta}_{ML,n}(\omega_1, \ldots, \omega_n)
:= \mathop{\rm argmax}_{\bm{\theta}} p^{k_1}_{\bm{\theta}} \cdots p^{k_n}_{\bm{\theta}}$, 
satisfies the equality ($\cong$) when $n$ is sufficiently large. Therefore, for parametric estimation in a family of probability distributions, the inverse $\bm{j}_{\bm{\theta}}^{-1}$ of the Fisher information matrix signifies the asymptotically optimal mean-square error. For its validity, some conditions on an estimator $M^n$ are needed for the inequality (\ref{kka}). This fact shows that the mean-square error matrix can be minimized asymptotically. These discussions with sufficiently large $n$ are called asymptotic theory, and its general and rigorous discussion has been established for classical cases. To discuss this problem more rigorously, however, we need to clarify the conditions for estimators \cite{vdv}.

Note that, in the quantum framework, when density operators are noncommutative, the asymptotically optimal mean square error matrix is not unique even in the asymptotic setting. This is because we have to consider a tradeoff between the mean-square errors of respective parameters. Thus, we take the strategy of minimizing the trace of the product of the mean-square error matrix and a weighted matrix.

\section{Estimation of the expectation parameter in the family of the multi-mode quantum Gaussian states}\Label{s9}
Here, we discuss the estimation only of the expectation parameter in the family of the quantum Gaussian states 
$\{\rho_{\bm{\theta},\bm{v}}| \theta \in \real^{2l}\}$ defined in \S\ref{s5}. 
Since the state of this system is transformed to $\rho_{\sqrt{n}\bm{\theta},\bm{v}} \otimes \rho_{0,\bm{v}}^{\otimes (n-1)}$ by an adequate unitary, the structure of the state family concerning the unknown parameter $\bm{\theta}$ is invariant even if the number $n$ of states increases. Therefore, we can apply the discussion for $n=1$ to the cases of $n > 2$.

Here, we will restrict our estimators to unbiased ones. Later, we will go on to the discussion of an arbitrary family of quantum states, in which for sufficiently large~$n$, we will treat estimators that may be biased.

Like (\ref{Cr}), the following right logarithmic derivative (RLD) Cram\'{e}r--Rao inequality holds:
\begin{align}
\bm{v}_{\bm{\theta}}(M) \ge \bm{j}_{\bm{\theta},R}^{-1} \Label{CrR},
\end{align}
where the RLD Fisher matrix $\bm{j}_{\bm{\theta},R}$ and the RLD 
$L_{l;\bm{\theta}}$ are defined as 
\begin{align*}
j_{k,l;\bm{\theta},R} := 
\Tr \rho_{\bm{\theta}} L_{l;\bm{\theta}} (L_{k;\bm{\theta}})^* ,\quad
\frac{\partial \rho_{\bm{\theta}}}{\partial \theta^l} 
 = \rho_{\bm{\theta}} L_{l;\bm{\theta}}.
\end{align*}
Like (\ref{Cr}), the Schwarz inequality plays an essential role in its proof. By calculation, for the family of the Gaussian states, the RLD Fisher matrix $\bm{j}_{\bm{\theta},R}$ has the following inverse matrix:
\begin{align} \bm{j}_{\bm{\theta},R}^{-1} = \bm{v}+ i \bm{s}. \Label{CrR2}
\end{align}
Here, taking the trace, we get
\begin{align}\tr \bm{v}_{\bm{\theta}}(M)
\ge \min\{ \tr \bm{a} |\bm{a} \hbox{ is symmetric, } \bm{a} 
\ge \bm{v}+ i \bm{s}\hbox{, and }\bm{a} \ge 0 \}=
\tr (\bm{v} + |\bm{s}|),
\end{align}
so we have a lower bound. On the other hand, by (\ref{822}), when the state is $\rho_{\bm{\theta}, \bm{v}}$, if we perform the measurement corresponding to a POVM $T_{|\bm{s}|+ \epsilon}(\bm{X}):= \{T_{\bm{\theta},|\bm{s}|+ \epsilon} (X)\}$, then the observed value is distributed as the normal distribution with the mean $\bm{\theta}$ and the variance $\bm{v}+|\bm{s}|+ \epsilon$.  
Therefore, we get
\begin{align}
\inf\{\tr \bm{v}_{\bm{\theta}}(M)| M\hbox{ is unbiased.}\}=
\tr (\bm{v} + |\bm{s}|)
\end{align}
(see \cite{HolP}). 
This formulation is generalized to the minimization of the weighted sum of the elements of $\bm{v}_{\bm{\theta}}(M)$, in which we minimize the error $\tr \bm{g}\bm{v}_{\bm{\theta}}(M)$ expressed with a weight matrix $\bm{g}\ge 0$. Note that there are many quantum extensions of logarithmic derivatives owing to noncommutativity, but this definition is the most suitable when we discuss only the family of the quantum Gaussian states. In a later section, we will explain its relation with another logarithmic derivative. For derivation of (\ref{CrR}) and (\ref{CrR2}), see Holevo~\cite{HolP} or Hayashi \cite{H-nyu}.

\section{Asymptotic theory for an arbitrary family of quantum states}\Label{s10}
In a similar way to classical systems, it is not easy to optimize our estimator with an arbitrary family of quantum states when the number of states $n$ is finite. Thus, we will discuss the case when $n$ is sufficiently large, that is, for the asymptotic case: We will treat a sequence of estimators $\vec{M}=\{M^n\}$ and we impose the following asymptotic local covariance condition at each point $\bm{\theta}$, instead of the unbiasedness. Thus, we define the following to represent a distribution of difference $\frac{\bm{x}}{\sqrt{n}}$ between the true value $\bm{\theta}$ and the estimate: 
\begin{align}
P^{M^n}_{\bm{\theta}}(B):= \Tr \rho_{\bm{\theta}}^{\otimes n} M^n \left\{ \left. \bm{\theta} + \frac{\bm{x}}{\sqrt{n}}\right|\bm{x} \in B \right\}. \end{align} 
Note that $P^{M^n}_{\rho^{\otimes n}_{\bm{\theta}}}(B)$ is different from $P^{M^n}_{\bm{\theta}}(B)$. If 
\begin{align}
\lim P^{M^n}_{\bm{\theta}}(B)= 
\lim P^{M^n}_{\bm{\theta}+ \frac{\bm{c}}{\sqrt{n}}}(B) ,\quad
\bm{v}_{\bm{\theta}}(\vec{M}) 
= \lim_{n \to \infty} n \bm{v}_{\bm{\theta}}(M^n)
\Label{kkd}
\end{align}
hold for each $\bm{c}\in\real^d$, then $\vec{M}$ is said to be asymptotically locally covariant at $\bm{\theta}$, and we will write
$ P^{\vec{M}}_{\bm{\theta}}(B)= 
\lim P^{M^n}_{\bm{\theta}}(B)$.
Then, $\bm{v}_{\bm{\theta}}(\vec{M})$ is expressed by 
\begin{align*}
v_{\bm{\theta}}^{k,j}(\vec{M})=
\int x^kx^j P_{\bm{\theta}}^{\vec{M}}(d x).
\end{align*}
Furthermore, if an estimator is asymptotically locally covariant at all $\bm{\theta}$, it is simply said to be asymptotically locally covariant.

In the following, we will treat the value 
\begin{align*}
C^A_{\bm{\theta}}(\bm{g}):= \inf\{
\tr \bm{v}_{\bm{\theta}}(\vec{M}) \bm{g} | \vec{M} \hbox{ is asymptotically locally covariant at }\bm{\theta}\},
\end{align*}
expressed with a weight matrix $\bm{g}\ge 0$. In particular, it is simplified to~$C^A_{\bm{\theta}}$ when $\bm{g}$ is the identity matrix.

To perform the measurement corresponding to an arbitrary POVM on ${\cal H}^{\otimes n}$, we need quantum interaction among systems, which is usually difficult. As a class of measurements implemented easily, we consider the following class illustrated below: 
Each measurement is performed for each quantum system but the selection of measurements is possible adaptively according to the data obtained in all measurements before: Depending on the first to $(k-1)$th observed values $(\omega_1, \ldots , \omega_{k-1}) \in \Omega^{k-1}$, the $k$th measurement $M^k(\omega_1, \ldots , \omega_{k-1})$ can be selected. This is the same setting as the classical adaptive experimental design, and a POVM corresponding to this measurement is said to be adaptive. 
\begin{align*}
\xymatrix{
 &\hbox{Measurement apparatus} & \hbox{Observed value} & 
\hbox{Estimate} \\
\rho_{\bm{\theta}} \ar@{|->}[r] & 
*+[F]{M^1}\ar[r]& \omega_1 
\ar@{|->}[ddr]\ar@{|->}[dl]\ar@{|->}[ddl]\ar@{|->}[ddddl]& \\
\rho_{\bm{\theta}} \ar@{|->}[r] & 
*+[F]{M^2}\ar[r] & \omega_2 
\ar@{|->}[dr]\ar@{|->}[dl]\ar@{|->}[dddl]& \\
\rho_{\bm{\theta}} \ar@{|->}[r] & 
*+[F]{M^3}\ar[r] & \omega_3 \ar@{|->}[r]\ar@{|->}[ddl]& 
\hat{\theta}_n (\omega_1, \cdots, \omega_n) \\
\vdots& \vdots & \vdots &\\
\rho_{\bm{\theta}} \ar@{|->}[r] & 
*+[F]{M^n}\ar[r] & \omega_n \ar@{|->}[uur]& \\
& &\hbox{\rm Figure 1} &
\save "2,1"."6,1"*[F.]\frm{} 
\restore
}
\end{align*}
An adaptive POVM can be written as a POVM on ${\cal H}^{\otimes n}$ as 
\begin{eqnarray}
\{M^1_{\omega_1}\otimes M^2(\omega_1)_{\omega_2}\otimes
M^3(\omega_1,\omega_2)_{\omega_3}\otimes
\cdots
\otimes M^n(\omega_1, \ldots , \omega_{n-1})_{\omega_n}
\}_{(\omega_1, \ldots , \omega_n) \in \Omega^n}.
\Label{kg1}
\end{eqnarray}
Furthermore, as a wider class of POVMs, we sometimes consider a class whose element $M(\omega)$ can be expressed as a tensor product. A POVM in this class is said to be separable and can be written~as
\begin{eqnarray}
\{M^1(\omega) \otimes M^2(\omega)
\otimes \cdots \otimes M^n(\omega) \}_{\omega}.
\end{eqnarray}
Like $C^A_{\bm{\theta}}(\bm{g})$, we note
\begin{align*}
C^{A,a}_{\bm{\theta}} (\bm{g}) &:= \inf\{
\tr \bm{v}_{\bm{\theta}}(\vec{M})\bm{g} | \vec{M} \hbox{ is adaptive and asymptotically locally covariant at }
\bm{\theta}\},
\\
C^{A,s}_{\bm{\theta}} (\bm{g})& := \inf\{
\tr \bm{v}_{\bm{\theta}}(\vec{M})\bm{g} | \vec{M}\hbox{ is separable and asymptotically locally covariant at }
\bm{\theta}\}.
\end{align*}
By the inclusion relation between these classes, we see $C^{A,a}_{\bm{\theta}} (\bm{g}) \ge C^{A,s}_{\bm{\theta}} (\bm{g})$, but they coincide as we will explain later. When $\bm{g}$ is the identity matrix, they are denoted as $C^{A,a}_{\bm{\theta}}$ and $C^{A,s}_{\bm{\theta}}$, respectively. 

Next, to discuss $C^A_{\bm{\theta}}(\bm{g})$ from different viewpoints, we consider local unbiasedness as a weaker condition than the unbiasedness condition: Approximating the unbiasedness condition in a neighborhood of $\bm{\theta}_0 \in \Theta$, we get
\begin{align}
e_{\bm{\theta}_0}^k(M)= \theta_0^k , \quad
\left . \frac{\partial e_{\bm{\theta}}^k(M)}{\partial \theta^j}\right|_{\bm{\theta}=\bm{\theta}_0}
= \delta^k_j.
\end{align}
Thus, this is called the local unbiasedness condition at $\bm{\theta}_0 \in \Theta$.

Based on the local unbiasedness condition, we define the following value:
\begin{align}C^1_{\bm{\theta}}(\bm{g})
:= \inf \{
\tr \bm{v}_{\bm{\theta}}(M)\bm{g}| M \hbox{ is locally unbiased at }\bm{\theta}\}.\end{align}
In particular, if $\bm{g}$ is the identity matrix, it is denoted as $C^1_{\bm{\theta}}$.

Generally, in the commutative case, that is, when we consider a family of probability distributions, $C^1_{\bm{\theta}}(\bm{g}) = \Tr \bm{j}_{\bm{\theta}}^{-1}\bm{g}$ holds: There exists a locally unbiased estimator whose covariance matrix is $\bm{j}_{\bm{\theta},S}^{-1}$. 
Conversely, the covariance matrix of any locally unbiased estimator is lower bounded by $\bm{j}_{\bm{\theta}}^{-1}$. 
Thus, for a POVM $M$, denoting the Fisher information matrix of a family of probability distributions by $\bm{j}_{\bm{\theta}}^M$, we get
\begin{align}
C^1_{\bm{\theta}}(\bm{g})
= \inf \{
\tr (\bm{j}_{\bm{\theta}}^M)^{-1} \bm{g}| M\hbox{: POVM}\}.
\end{align}

In addition, let $C^n_{\bm{\theta}}(\bm{g})$ denote the value $C^1_{\bm{\theta}}(\bm{g})$ of the family of density operators $\{\rho_{\bm{\theta}}^{\otimes n}| \bm{\theta} \in \Theta\}$. Comparing their selections of measurements, we can easily show $C^1_{\bm{\theta}}(\bm{g}) \ge n C^n_{\bm{\theta}}(\bm{g})$. The following theorem holds: 
\begin{thm}\Label{t3} 
By letting $\bm{v}_{\bm{\theta}}(\bm{X}):=\bm{v}_{\rho_{\bm{\theta}}}(\bm{X})$ and $\bm{s}_{\bm{\theta}}(\bm{X}):=\bm{s}_{\rho_{\bm{\theta}}}(\bm{X})$, the following equalities hold: 
\begin{align}
C^A_{\bm{\theta}}(\bm{g})& = \lim_{n \to \infty} n C^n_{\bm{\theta}}(\bm{g}) 
=C^H_{\bm{\theta}}(\bm{g}):=
\min_{\bm{X}}\left\{ \tr \left(
\bm{v}_{\bm{\theta}}(\bm{X})\bm{g} +
 |\sqrt{\bm{g}}\bm{s}_{\bm{\theta}}(\bm{X})\sqrt{\bm{g}}| \right)\left|
\Tr X^k \frac{\partial \rho_{\bm{\theta}}}{\partial \theta^l}=\delta^k_l\right.\right\},
\Label{4-26-1}
\\
C^{A,a}_{\bm{\theta}}(\bm{g})&=C^{A,s}_{\bm{\theta}}(\bm{g})= C^1_{\bm{\theta}}(\bm{g}).
\Label{4-26-2}
\end{align}
Furthermore, the bounds $C^A_{\bm{\theta}}(\bm{g})$ and $C^{A,a}_{\bm{\theta}}(\bm{g})$ can be attained at all points~$\bm{\theta}$ by a single estimator in the respective classes. 
\end{thm}

From this theorem, we can consider that $C^1_{\bm{\theta}}(\bm{g})$ and $C^H_{\bm{\theta}}(\bm{g})$ represent the bounds of accuracy of quantum estimation. The former represents the bound with no quantum correlation in measuring apparatus, while the latter represents the bound when the quantum correlation is positively used in measuring apparatus. 
It is known that they coincide when $d=1$ (Helstrom~\cite{Hel}) or when $\rho_{\bm{\theta}}$ is the pure state (Matsumoto~\cite{Matu-Memo}).

To analyze these values further, we introduce the symmetric logarithmic derivative (SLD) $L_{k;\bm{\theta}}$ and the SLD Fisher information matrix $\bm{j}_{\bm{\theta},S}$ as follows:
\begin{align}
\frac{\partial \rho_{\bm{\theta}}}{\partial \theta^k}:=
L_{k;\bm{\theta}} \circ \rho_{\bm{\theta}}, \quad
j_{k,l;\bm{\theta},S}
:= \Tr \rho_{\bm{\theta}} (L_{k;\bm{\theta}} \circ L_{l;\bm{\theta}}).
\end{align}
When $d=1$, we can show $C^H_{\bm{\theta}}(\bm{g})= \bm{j}_{\bm{\theta},S}^{-1}= C^1_{\bm{\theta}}(\bm{g})$. Here, we define a superoperator $D_{\bm{\theta}}$ as what satisfies $\Tr (D_{\bm{\theta}}(X)\circ Y) \rho_{\bm{\theta}}= -i \Tr [X,Y]\rho_{\bm{\theta}}$. If all operators $D_{\bm{\theta}}(L_{\bm{\theta}}^1) ,\ldots, D_{\bm{\theta}}(L_{\bm{\theta}}^d)$ can be expressed as a linear combination of $L_{\bm{\theta}}^1 ,\ldots, L_{\bm{\theta}}^d$, then
\begin{align}C^H_{\bm{\theta}}(\bm{g})= 
\tr \left(
\bm{v}_{\bm{\theta}}(\bm{L}^{-1}_{\bm{\theta}}) \bm{g} +
| \sqrt{\bm{g}} \bm{s}_{\bm{\theta}}(\bm{L}^{-1}_{\bm{\theta}}) \sqrt{\bm{g}}|
\right),\end{align}
where we define $L^{-1,k}_{\bm{\theta}}:= \sum_l (j_{\bm{\theta},S}^{-1})^{k,l} L_{l;\bm{\theta}}$. Then,
\begin{align}
\bm{v}_{\bm{\theta}}(\bm{L}^{-1}_{\bm{\theta}}) +
\bm{s}_{\bm{\theta}}(\bm{L}^{-1}_{\bm{\theta}})=
\bm{j}_{\bm{\theta},R}^{-1}
\end{align}
holds. In particular, for the family of the quantum Gaussian states, this condition holds and we can check that $\Tr ( \bm{v} \bm{g}+|\sqrt{\bm{g}}\bm{s} \sqrt{\bm{g}}|)$ is equal to the bound $C^H_{\bm{\theta}}(\bm{g})$. 
Then, there is no difference between $C^1_{\bm{\theta}}(\bm{g})$ and $C^H_{\bm{\theta}}(\bm{g})$. 
Furthermore, 
if $\bm{s}_{\bm{\theta}}(\bm{L}^{-1}_{\bm{\theta}})=0$ holds, 
then we can show
\begin{align}
C^H_{\bm{\theta}}(\bm{g})= 
\tr \bm{g} \bm{j}_{\bm{\theta},S}^{-1}.
\end{align}

On the other hand, Gill and Massar \cite{GM} proved the inequality
\begin{align}
\tr \bm{j}_{\bm{\theta},S}^{-1} \bm{j}_{\bm{\theta}}^M \le 
\dim {\cal H} -1.
\end{align}
In particular, for $\dim {\cal H}=2$, the right-hand side is 1. 
When $\bm{j}_{\bm{\theta},S}^{-\frac{1}{2}} 
\bm{j}_{\bm{\theta}}^M
\bm{j}_{\bm{\theta},S}^{-\frac{1}{2}} $ is a one-dimensional projection,
the measurement $M$ can be realized by the spectral decomposition of 
the Hermitian matrix 
that is the eigenvector with the eigenvalue 1 of 
$\bm{j}_{\bm{\theta},S}^{-\frac{1}{2}} 
\bm{j}_{\bm{\theta}}^M
\bm{j}_{\bm{\theta},S}^{-\frac{1}{2}}$.
Hence, when $\tr \bm{j}_{\bm{\theta},S}^{-1} \bm{j}_{\bm{\theta}}^M=1$, this measurement can be constructed by a probabilistic combination of the measurements above.
Thus, the set of all realizable measurements is 
$\{M| \tr \bm{j}_{\bm{\theta},S}^{-1} \bm{j}_{\bm{\theta}}^M =1\}$.
Using Lagrange multiplier method,
we obtain
\begin{align}
C^1_{\bm{\theta}}(\bm{g})= \left(\tr \sqrt{\bm{j}_{\bm{\theta},S}^{-1/2}
\bm{g} \bm{j}_{\bm{\theta},S}^{-1/2}}\right)^2.\Label{hs}
\end{align}
For two parameters, Nagaoka~\cite{Nagaoka:1991} proved (\ref{hs}), while for three parameters Hayashi \cite{Haya1,Haya1-2} proved it first using the duality theorem of linear programming. We can further prove the following lemma:
\begin{lem}\Label{l1}
When $\dim {\cal H}=2$, for $\bm{g} \,> 0$,
\begin{align}
C^1_{\bm{\theta}}(\bm{g})=C^H_{\bm{\theta}}(\bm{g}) \Label{l1-1}
\end{align}
holds if and only if $d=1$ holds or $d=2$ and $\rho_{\bm{\theta}}$ is a pure state.
\end{lem}

An outline of the proof of Theorem \ref{t3}: 
We will prove the theorem by showing some inequalities. 
The most difficult part $\lim n C^n_{\bm{\theta}}(\bm{g}) \le C^H_{\bm{\theta}}(\bm{g}) $ will be proved in a later section because of not only its difficulty but also its connection with the quantum central limit theorem. 

First, we will prove 
$C^{A,s}_{\bm{\theta}}(\bm{g}) \le C^1_{\bm{\theta}}(\bm{g}) $. 
By selecting a POVM $M_{\bm{\theta}}$ satisfying
$\tr (\bm{j}_{\bm{\theta}}^{M_{\bm{\theta}}})^{-1} \bm{g} 
= C^1_{\bm{\theta}}(\bm{g})$, the bound can be attained at $\bm{\theta}$. 
From this fact, we will show
$C^{A,a}_{\bm{\theta}}(\bm{g}) \le C^1_{\bm{\theta}}(\bm{g})$
as follows: We measure the first $\sqrt{n}$ states with a POVM $M'$ 
satisfying $\bm{j}_{\bm{\theta}}^{M'} \,> 0$ for all $\bm{\theta}$, and we estimate the unknown parameter $\bm{\theta}$ by the maximum likelihood estimator $\hat{\bm{\theta}}$ based on $\sqrt{n}$ data. 
We measure each of the remaining $n- \sqrt{n}$ systems by the measurement $M_{\hat{\bm{\theta}}}$. 
Then, the bound $C^1_{\bm{\theta}}(\bm{g})$ can be attained uniformly for all $\bm{\theta}$ \cite{GM,HM}. 
This shows that the bound $C^1_{\bm{\theta}}(\bm{g})$ can be uniformly attained by an adaptive POVM\null. 
Applying this argument to the system ${\cal H}^{\otimes m}$, we can prove $C^A_{\bm{\theta}}(\bm{g}) \le mC^m_{\bm{\theta}}(\bm{g})$ and we can show that $mC^m_{\bm{\theta}}(\bm{g})$ can be attained for all $\bm{\theta}$. Then, letting $m \to \infty$, we get $C^A_{\bm{\theta}}(\bm{g}) \le
\lim_{m \to \infty} m C^m_{\bm{\theta}}(\bm{g}) $.

From the above, we obtain (\ref{4-26-1}) if 
$n C^n_{\bm{\theta}}(\bm{g}) \ge C^H_{\bm{\theta}}(\bm{g}) $ and
$C^A_{\bm{\theta}}(\bm{g}) \ge C^H_{\bm{\theta}}(\bm{g}) $
are proved. It requires several pages to prove these inequalities rigorously, so we will write only an outline. 
Letting $X^k:= \int_{\real^d} \hat{\theta}^k M^n(\,d \hat{\bm{\theta}})$, we get the former for $n=1$. 
Then, the former inequality for arbitrary $n$ follows from the fact that
the vector $\bm{X}$ minimizing 
$\tr \left(\bm{v}_{\bm{\theta}}(\bm{X})\bm{g} +
 |\sqrt{\bm{g}}\bm{s}_{\bm{\theta}}(\bm{X})\sqrt{\bm{g}}| \right)$
in (\ref{4-26-1}) belongs to the range of the superoperator $D_{\bm{\theta}}$ whose domain is the linear space spanned by $L_1, \ldots, L_d$.
The latter is proved as follows:
First, the asymptotic local unbiasedness concerning $\vec{M}$ is defined by extending the local unbiasedness.
Then, we prove that the asymptotic local unbiasedness at $\bm{\theta}$ implies the asymptotic local covariance at $\bm{\theta}$. 
An inequality similar to the Cram\'{e}r--Rao inequality is derived by using the Schwarz inequality. Similarly, we can show $C^{A,s}_{\bm{\theta}}(\bm{g}) \ge C^1_{\bm{\theta}}(\bm{g}) $. Therefore, we obtain (\ref{4-26-2}). 

We will explain the condition of local unbiasedness a little. 
In fact, it has a clear meaning to minimize the mean-square error under the local unbiasedness condition while this condition is not so natural. This is because this minimum mean-square error coincides with the minimum mean-square error under the more natural condition, i.e., the asymptotically locally covariance. This causes from the relation between the minimum mean-square error and the Fisher information matrix $\bm{j}^M_{\bm{\theta}}$ for a fixed measurement $M$. However, if we adopt an error different from the mean-square error---for example, the absolute value of the difference from the true parameter or its $s(\,< 2)$th power as a loss function---then such meaning cannot hold. 
In particular, if $\rho_\theta$ is a pure state, then the infimum of such an error is $0$ under the local unbiasedness condition \cite{Haya1-2}. Therefore, there is no correspondence of this infimum with the error of the optimal covariant estimator given in Hayashi~\cite{H-d}, which is the optimum concerning the family of all pure states in a finite-dimensional system under the covariant or the minimax criterion. 

\section{Estimation with quantum correlation of the family of the one-mode quantum Gaussian states}\Label{s11}
We will discuss the simultaneous estimation of $\zeta$ and $N$ in the family of the one-mode quantum Gaussian states $\{\rho_{\zeta,N}| \zeta \in \complex, N\,> 0 \}$ given in \S\ref{s4}. In this model, the optimal measurement can be physically realized and a lower bound can be explicitly obtained. Hence, we can understand ``How does the quantum correlation effect the quantum estimation?'' 
When the size of noise $N$ is known, as we discussed in \S\ref{s9},
\begin{eqnarray*}
C^{A,s}_{\zeta}= C^{A}_{\zeta}= 2 (N+1) \end{eqnarray*}
holds, so there is no advantage to use quantum correlation. When we perform the heterodyne measurement (\ref{10}), we can construct an estimator attaining this lower bound \cite{HolP}.

Now, consider estimating not only $\zeta$ but also $N$. 
Then, the relation
\begin{eqnarray*}
C^{A,s}_{\zeta,N} \,> C^{A}_{\zeta,N}= (N+2) (N+1)
= 2 (N+1)+ N(N+1)
\end{eqnarray*}
holds for the family of density operators $ {\cal S}^g:= \{ \rho_{\zeta,N} | \zeta \in \complex, N \,> 0\}$, so we obtain an effect to use quantum correlation \cite{Hayashi:2000,bussei}.

Regrettably, the value of $C^{A,s}_{\zeta,N}$ has not been obtained concretely. Here, using a different method, we will evaluate the difference between presence and absence of quantum correlation. Restricting the sum of the mean-square errors for a complex parameter $\zeta (= \zeta_1+ \zeta_2 i)$ to be $\frac{2 (N+1)}{n}$, we will focus on the minimum mean-square error concerning the other parameter $N$. This minimum value (per one prepared state) is $(N+1)^2$ without quantum correlation, while it is asymptotically equal to $N(N+1)$ with quantum correlation. In this setting, the heterodyne measurement (\ref{10}) is the optimum for the former, while the measurement in the following procedure is the optimum for the latter:

First, using an adequate interaction among $n$ systems, we will transform the system as 
\begin{equation*}
\rho_{\zeta,N}^{\otimes n} \mapsto
\rho_{\sqrt{n} \zeta,N} \otimes \rho_{0,N}^{\otimes n-1} .
\end{equation*}
Then, we perform the heterodyne measurement (\ref{10}) for the first system, and estimate the parameter $\xi$ by the observed value. We measure the photon numbers (\ref{9}) for the other $n-1$ systems, and estimate the other parameter $N$ by its mean. For example, for $n=4$, the proposed measurement can be realized by combining half mirrors as the following Figure~2: 
\begin{equation*}
\xymatrix{
&& &*+[F-]{{\bf N}} &&\\
\rho_{\zeta,N} & & \rho_{0,N}\ar@{.>}[ur] & & & \\
& *+[F-]{{\rm HM} }\ar@{.>}[dr]\ar@{.>}[ur] \ar@{<.}[dl] \ar@{<.}[ul] 
& & &  & *+[F-]{{\bf N}}
\\
\rho_{\zeta,N} & & \rho_{\sqrt{2}\zeta,N} &&\rho_{0,N}\ar@{.>}[ur] & \\
& & & *+[F-]{{\rm HM} }\ar@{.>}[dr]\ar@{.>}[ur] \ar@{<.}[dl]
\ar@{<.}[ul] & & \\
\rho_{\zeta,N} & & \rho_{\sqrt{2}\zeta,N} &&\rho_{2\zeta,N}\ar@{.>}[dr]& \\
& *+[F-]{\rotatebox[origin=c]{180}{\rm HM} } 
\ar@{.>}[dr]\ar@{.>}[ur] \ar@{<.}[dl] \ar@{<.}[ul] &&& & *+[F-]{{\bf H}}\\
\rho_{\zeta,N} & & \rho_{0,N}\ar@{.>}[dr] && & \\
&& &*+[F-]{{\bf N}} &&\\
 & &\ar@{}[r]|{\hbox{\rm Figure 2}} & & & 
\save"2,1"."8,1"*[F.] 
\frm{} 
\restore}
\end{equation*}
where $\xymatrix{*+[F-]{{\bf N}}}$ and $\xymatrix{*+[F-]{{\bf H}}}$ respectively signify number measurement (\ref{9}) and heterodyne measurement (\ref{10}), while $\xymatrix{*+[F-]{{\rm HM}}}$ signifies a half mirror: precisely, $\xymatrix{*+[F-]{\rotatebox[origin=c]{180}{\rm HM}}}$ is 
180-degree rotation of a half mirror $\xymatrix{*+[F-]{{\rm HM}}}$.

\section{The quantum central limit theorem and the bound of the estimation error}\Label{s12}
Using the quantum central limit theorem, we will construct a sequence of locally unbiased estimators that asymptotically attains $C^H_{\bm{\theta}}(\bm{g}) $. 
First, we select a sequence of operators $\bm{X}=(X^1, \ldots, X^d)$ on ${\cal H}$ satisfying
\begin{align*}
C^H_{\bm{\theta}}(\bm{g}) = \tr \left(\bm{v}_{\bm{\theta}}(\bm{X})\bm{g} +
 |\sqrt{\bm{g}}\bm{s}_{\bm{\theta}}(\bm{X})\sqrt{\bm{g}}|\right) , \quad
\Tr X^k \frac{\partial \rho_{\bm{\theta}}}{\partial \theta^j}=\delta_j^k
\end{align*}
and define an operator $S_{n,R,\bm{v}'}$ on ${\cal H}^{\otimes n}$ as
\begin{align}
S_{n,R,\bm{v}'}:=
\int_{\|\bm{x}\| \le R}
T_{\bm{x},\bm{v}'}(\bm{X}^{(n)}) \,d \bm{x}.
\end{align}
Then, we define a POVM 
$M^n= \{M^n( \hat{\bm{\theta}}) \}_{\|\hat{\bm{\theta}}\|\le R/\sqrt{n}}$ by
\begin{align}
M^n_{\bm{v}',\bm{X},R}
( \hat{\bm{\theta}}):=
\sqrt{n}
S_{n,R,\bm{v}'}^{-1/2}
T_{\sqrt{n} \hat{\bm{\theta}},\bm{v}'}
(\bm{X}^{(n)}) S_{n,R,\bm{v}'}^{-1/2}.
\end{align}
Here, as we will discuss later, for $\bm{v}'$ satisfying 
$\bm{v}' + i \bm{s}_{\bm{\theta}} (\bm{X}) \ge 0$, 
there is a sequence $\{R_n(\,> 0)\}$ such that 
\begin{align}
\lim_{n \to \infty} \frac{\partial e^k_{\bm{\theta}}
(M^n_{\bm{v}',\bm{X},R_n})}
{\partial \theta^j}= \delta^k_j ,\quad 
\lim_{n \to \infty} n v^{k,j}_{\bm{\theta}}(M^n_{\bm{v}',\bm{X},R_n})
= v^{k,j}_{\bm{\theta}}(\bm{X}) + {v'}^{k,j} \Label{20-1}.
\end{align}
Now, we define $A_{j;n}^k:= \frac{\partial e^k_{\bm{\theta}}(M^n_{\bm{v}',\bm{X},R_n})}
{\partial \theta^j}$,
and we introduce another outcome 
$\hat{\bm{\theta}}':= \bm{A}_n^{-1}\hat{\bm{\theta}}$.
The estimator $M^n_{\bm{v}',\bm{X},R_n}$ with the outcome $\hat{\bm{\theta}}'$ is locally unbiased. Its covariance matrix is 
$\bm{A}_n^{-1} \bm{v}_{\bm{\theta}}(M^n_{\bm{v}',\bm{X},R_n}) (\bm{A}_n^{-1})^*$ and it converges to $\bm{v}_{\bm{\theta}}(\bm{X}) + \bm{v}'$.

Now, we select $\bm{v}'$ by $\bm{v}'= \sqrt{\bm{g}}^{-1} \left( |\sqrt{\bm{g}}\bm{s}_{\bm{\theta}}(\bm{X})\sqrt{\bm{g}}| + \epsilon \right)\sqrt{\bm{g}}^{-1} $ for any $\epsilon \,> 0$. Then, by taking the limit $\epsilon \to 0$, it attains $C^H_{\bm{\theta}}(\bm{g})$. Here, we cannot substitute $0$ into $\epsilon$ directly. This is because there is an eigenvalue $1$ in the matrix $|(\bm{v'})^{-\frac{1}{2}} \bm{s} (\bm{v'})^{-\frac{1}{2}}|$, which is the variable of $f$ in the definition of $T_{\sqrt{n} \hat{\bm{\theta}},\bm{v}'}$ at (\ref{4-25-1}). In fact, it diverges when the variable of $f$ is $1$. 

We will write an outline of the proof of (\ref{20-1}). Essentially, (\ref{20-1}) follows from Theorem~\ref{t2}, but treatment of $S_{n,R,\bm{v}'}^{-1/2} $ is somewhat difficult. We need to show that this operator converges to $I$ in some sense asymptotically in a neighborhood of $\rho_{\bm{\theta}}^{\otimes n}$. To show this, in this context, we can check that it is enough to prove that $\Tr \rho_{\bm{\theta}}^{\otimes n} (I-S_{n,R,\bm{v}'} )^2$ converges to $0$ by technical treatments of inequalities. This convergence follows from Theorem~\ref{t2} and its small extension.

The point of this proof is to derive the convergence with respect to probability distributions from the central limit theorem guaranteeing the convergence with respect to algebraic structure (Theorem~\ref{t1}). Since it was easier to extend the central limit theorem to quantum systems based on the algebraic structure, Theorem~\ref{t1} was obtained first. However, the original central limit theorem guarantees that the distribution converges to the normal one, so its quantum extension should guarantee the convergence with respect to distributions in some sense. The argument presented in this section says that simultaneous distributions by a quantum measurement converge to the simultaneous distribution by the corresponding quantum measurement under the family of the quantum Gaussian states. 
This can be called a quantum version of the central limit theorem satisfying the condition above. However, since the proof is complicated and is not straightforward enough, it is understood that there is a possibility for improvement. There might be a quantum central limit theorem in a more natural form. Attainment of the asymptotic bound described here might follow from such a theorem and we can expect development toward this direction. 


\section{Future prospects}\Label{s13}
In this article, to construct an estimator that works uniformly efficiently for an arbitrary family of quantum states, we featured a two-stage method of estimation: First, we estimate the parameter with $\sqrt{n}$ states, and we make measurement working locally efficiently for the rest of the states. Since such an estimator is not so natural, we hope that a more natural estimator will be constructed. An estimator $\{M^n(\hat{\bm{\theta}})\}_{\hat{\bm{\theta}}}$ that constructed below is expected to work uniformly efficiently, though its asymptotic performance has not be analyzed enough. 
For $\{\bm{g}_{\bm{\theta}}\}_{\bm{\theta}}$, we define 
\begin{align}
\bm{X}_{\bm{\theta}}&:= 
\argmin_{\bm{X}}
\left\{ \tr \left(
\bm{v}_{\bm{\theta}}(\bm{X})\bm{g}_{\bm{\theta}} +
 |\sqrt{\bm{g}_{\bm{\theta}}}\bm{s}_{\bm{\theta}}(\bm{X})
\sqrt{\bm{g}_{\bm{\theta}}}| \right)\left|
\Tr X^k \frac{\partial \rho_{\bm{\theta}}}{\partial \theta^j}=\delta^j_i
\right.\right\}, \nonumber \\
S_{n,\bm{v}_{\bm{\theta}}}&:=
\int_{\Theta}
T_{\bm{x},\bm{v}'_{\bm{\theta}}}(\bm{X}_{\bm{\theta}}^{(n)}) 
\,d \bm{\theta}, \nonumber \\
M^n(\hat{\bm{\theta}})&:=
S_{n,\bm{v}_{\hat{\bm{\theta}}}}^{-1/2}
T_{\hat{\bm{\theta}},\bm{v}'_{\hat{\bm{\theta}}}}
(\bm{X}_{\hat{\bm{\theta}}}^{(n)}) S_{n,\bm{v}_{\hat{\bm{\theta}}}}^{-1/2}.
\label{57}
\end{align}
Then, we conjecture that
\begin{align}
\lim_{n \to \infty} n \bm{v}_{\bm{\theta}}(M^n)
= \bm{v}_{\bm{\theta}}(\bm{X}) + \bm{v}'_{\bm{\theta}}
\Label{yoso}\end{align}
holds. If we exchange the orders of the limits without checking its validity, we can derive (\ref{yoso}) similarly to the argument in \S\ref{s12}. Of course, (\ref{yoso}) is only a conjecture because this derivation contains many operations that are not mathematically allowed. Therefore, we can expect that the bound would be attained by the estimator (\ref{57}) with $\bm{v}'_{\bm{\theta}}:= \sqrt{\bm{g}_{\bm{\theta}}}^{-1} \left(
|\sqrt{\bm{g}_{\bm{\theta}}}\bm{s}_{\bm{\theta}}(\bm{X}_{\bm{\theta}})\sqrt{\bm{g}_{\bm{\theta}}}| 
+ \epsilon \right)\sqrt{\bm{g}_{\bm{\theta}}}^{-1}$
in the limit $\epsilon \to 0$.

Furthermore, in classical systems, it is known that geometrical quantities such as curvatures play important roles in the asymptotic theory concerning higher-order errors, and the optimum higher-order errors can be geometrically characterized \cite{Amari,AN}. Similarly, we can expect that higher error terms would be related to geometrical quantities in some sense in the quantum case. 
Hayashi~\cite{H-kadai} treated higher-order asymptotic theory only in restricted models and pointed out that higher-order errors are related to a kind of geometrical quantity in these models. We hope that these facts will be analyzed from a more general viewpoint.

In future, to develop studies in this field, knowledge of classical mathematical statistics is of course needed, but also mathematical technique peculiar to quantum systems is often needed as we mentioned in \S\ref{s2}. Therefore, we need to grasp statistical inference of quantum systems not only from the closed viewpoint of mathematical statistics, but also from wider viewpoints including general information processing of quantum systems. Studies considering application to actual experiments are also needed.

Finally, the author thanks the referee for useful comments.

\subsection*{Additional note}
After the publication of the original Japanese version, two books treating quantum estimation were published \cite{selected,H-nyu}. The book \cite{selected} consists of the preceding papers after Holevo's famous book\cite{HolP}. It contains many references of this paper. In particular, it includes English translations of several important manuscripts originally written in Japanese.
The other book \cite{H-nyu} is a textbook covering quantum estimation and quantum information. Also one article \cite{HayaMa2} concerning quantum estimation was written. It rigorously treats the asymptotic quantum estimation in a quantum two level system (i.e., qubit system) and its relation to the estimation of the one-mode quantum Gaussian case.


\end{document}